\documentclass[%
 reprint,
 amsmath,amssymb,
 aps,
]{revtex4-2}

\usepackage{graphicx}% Include figure files
\usepackage{dcolumn}% Align table columns on decimal point
\usepackage{bm}% bold math
\begin{document}

\preprint{APS/123-QED}

\title{Tuning the interlayer coupling in \\ La$_{0.7}$Sr$_{0.3}$Mn$_{0.95}$Ru$_{0.05}$O$_3$ $/$ LaNiO$_3$  multilayers with perpendicular magnetic anisotropy\\}

\author{Jörg Schöpf}
 \email{schoepf@ph2.uni-koeln.de}
 \affiliation{II. Physics Institute, University of Cologne, 50937 Cologne, Germany}
 \author{Valentina Piva}
 \affiliation{II. Physics Institute, University of Cologne, 50937 Cologne, Germany}

 \author{Padraic Shafar}
 \affiliation{Advanced Light Source, Lawrence Berkeley National Laboratory, Berkeley, California 94720, USA}
\author{Divine P. Kumah}
\affiliation{Department of Physics, Duke University, Durham, NC 27517, USA and  Department of Physics, North Carolina State University, Raleigh, NC, 27695, USA} 

\author{Xuanyi Zhang}
\affiliation{Department of Physics, Duke University, Durham, NC 27517, USA  and Department of Physics, North Carolina State University, Raleigh, NC, 27695, USA}

 \author{Lide Yao}
 \affiliation{OtaNano-Nanomicroscopy Center, Aalto University, P.O. Box 15100, FI-00076 Aalto, Finland}

 \author{Sebastiaan van Dijken}
 \affiliation{Department of Applied Physics, Aalto University School of Science, P.O. Box 15100, FI-00076 Aalto}
\author{Paul H.M. van Loosdrecht}
\affiliation{II. Physics Institute, University of Cologne, 50937 Cologne, Germany}
\author{Ionela Lindfors-Vrejoiu}
 \email{vrejoiu@ph2.uni-koeln.de}
\affiliation{II. Physics Institute, University of Cologne, 50937 Cologne, Germany}


\date{\today}% It is always \today, today,

\begin{abstract}
In ferromagnetic oxide epitaxial multilayers, magnetic properties and interlayer coupling are determined by a variety of factors. Beyond the contribution of interlayer exchange coupling, strain and interfacial effects, such as structural reconstructions or charge transfer, play significant roles, resulting in complex magnetic behaviour. In this study the interlayer coupling of ferromagnetic La$_{0.7}$Sr$_{0.3}$Mn$_{0.95}$Ru$_{0.05}$O$_3$ (LSMRO) layers (8 nm thick) was investigated, when separated by epitaxial spacers of paramagnetic metallic LaNiO$_3$ (LNO), in  stacks exhibiting perpendicular magnetic anisotropy. By varying the thickness of the spacer, it was found that the coupling between two LSMRO layers changes from antiferromagnetic (with a 4 unit cell thick LNO spacer) to ferromagnetic (with a 6 unit cells thick LNO spacer). For multilayers comprising five LSMRO layers and a 4 unit cell thick LNO spacer, the antiferromagnetic coupling was preserved. However, the effective magnetic anisotropy changed, causing the magnetization to cant more towards the in-plane direction. This behavior was corroborated by X-ray magnetic circular dichroism (XMCD) investigations at the Mn and Ni L$_3$-edges. The XMCD results indicated that the 4 unit cells thick LNO spacer in the multilayer become magnetically ordered, closely following the magnetization of adjacent LSMRO layers.

%\begin{description}
%\end{description}
\end{abstract}

\maketitle

%\tableofcontents

\section{\label{sec:level1}Introduction}

Ferromagnetic oxides, especially the perovskite mixed-valance manganites with ordering temperatures exceeding room temperature, are a propitious material class for epitaxial heterostructures research \cite{Ijiri2002, Bhattacharya2014}. Perovskite manganites exhibit a delicate balance among competing interactions involving lattice, electronic, and spin  degrees of freedom \cite{Doerr2006}. Recent studies have revealed, complex non-collinear magnetic order in La$_{0.67}$Sr$_{0.33}$MnO$_3$-LaNiO$_3$ superlattices with in-plane magnetic anisotropy \cite{Hoffman2016, Fabbris2018}, as well as magnetic skyrmions bubbles in La$_{0.7}$Sr$_{0.3}$Mn$_{1-x}$Ru$_x$O$_3$ multilayers with perpendicular magnetic anisotropy \cite{Schoepf2023b}.

In magnetic multilayers, the type and the strength of interlayer coupling \cite{Stiles1993, Stiles1999} are crucial for determining their properties and realizing unique effects that are useful for applications, such as the giant magnetoresistance (GMR) \cite{Baibich1988}. For magnetic oxide epitaxial multilayers, the exploration of metallic non-magnetic oxides as spacer layers, enabling control over the type of coupling and compatible with coherent growth, has been less extensive compared to simpler ferromagnetic multilayers developed during the discovery of GMR. Recently findings demonstrated that for SrRuO$_3$ epitaxial layers with perpendicular magnetic anisotropy (PMA), the interlayer coupling is strongly ferromagnetic with 4 unit cells thick LaNiO$_3$ spacers \cite{Yang2021}, whereas SrIrO$_3$ spacer layers keep the SrRuO$_3$ decoupled \cite{Wysocki2022}. The use of paramagnetic metallic LaNiO$_3$ (LNO) spacer layers has been reported to result in oscillatory exchange coupling via the Ruderman–Kittel–Kasuya–Yosida (\textit{RKKY}) interaction in La$_{0.67}$Ba$_{0.33}$MnO$_3$-LNO heterostructures with in-plane magnetic anisotropy \cite{Nikolaev2000, Ohsawa2005}. However, given the low thickness of the LNO spacers used in these multilayers, the additional role of tunneling cannot be entirely ruled out.\\

In this work, we investigated the interlayer coupling between La$_{0.7}$Sr$_{0.3}$Mn$_{0.95}$Ru$_{0.05}$O$_3$ (LSMRO) ferromagentic layers with PMA, mediated by a LNO spacer layer. The PMA of the LSMRO layers arises from the modification of anisotropy due to Ru substitution of Mn, and the compressive strain from epitaxial growth on (LaAlO$_3$)$_{0.3}$-(Sr$_2$TaAlO$_6$)$_{0.7}$ (LSAT) substrates  \cite{Nakamura2018, Hua2022, Schoepf2023a, phdthesis_Lena}. We demonstrated that LNO is a suitable spacer material for LSMRO layers, enabling the tailoring of the interlayer coupling from antiferomagnetic (AFM) to ferromagnetic (FM) by adjusting the LNO thickness, likely through a RKKY-type mechanism \cite{Nikolaev2000, Ohsawa2005}. Controlling the coupling type between LSMRO layers enables the design of multilayers in which magnetic domains can couple either ferromagnetically or antiferromagnetically. Antiferromagnetic coupling in ferromagnetic oxide heterostructures is less commonly reported, although it holds significant promise, especially in the emerging field of antiferromagnetic spintronics \cite{Jungwirth2016} and memory devices\cite{Saito2021}. Synthetic antiferromagnets, consisting of two or more magnetic layers separated by spacers and coupled antiferromagnetically are envisioned for the development faster, smaller, more energy-efficient, and robust heterostructures that can enhance the information encoding in ultrahigh density spin-transfer-torque magnetic random access memory cells \cite{Wang2023}.  Furthermore, skyrmion bubbles in synthetic antiferromagnetic multilayers benefit from reduced skyrmion Hall-effect perturbations in their current-induced motion \cite{Dohi2019}. As the generation and density of skyrmions are strongly dependent on interlayer exchange coupling and easy-axis orientation \cite{Zhang2020}, understanding the strength of the coupling and how it changes with modifications in effective magnetic anisotropy in multilayers, is of paramount importance.

\begin{figure}[t]
    \centering
    \includegraphics[width=0.48\textwidth]{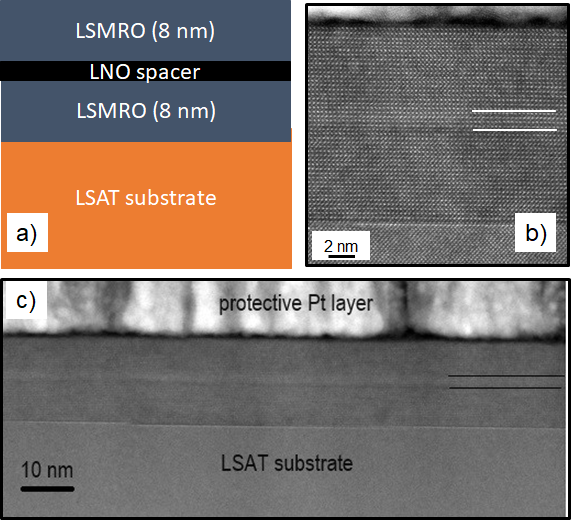}
    \caption{Structure of the samples used for the investigation of interlayer coupling. (a) Schematic drawing of the trilayer samples. (b) High magnification HAADF-STEM image of the cross-section of a trilayer sample with a 4 uc thick LNO spacer, marked by the white horizontal lines. (c) low magnification overview  HAADF-STEM image of the same sample, with the two black lines delineating the LNO spacer (The Pt protective layer was applied for STEM specimen preparation).}  
    \label{fig1}
\end{figure}

\begin{figure}[t]
    \centering
    \includegraphics{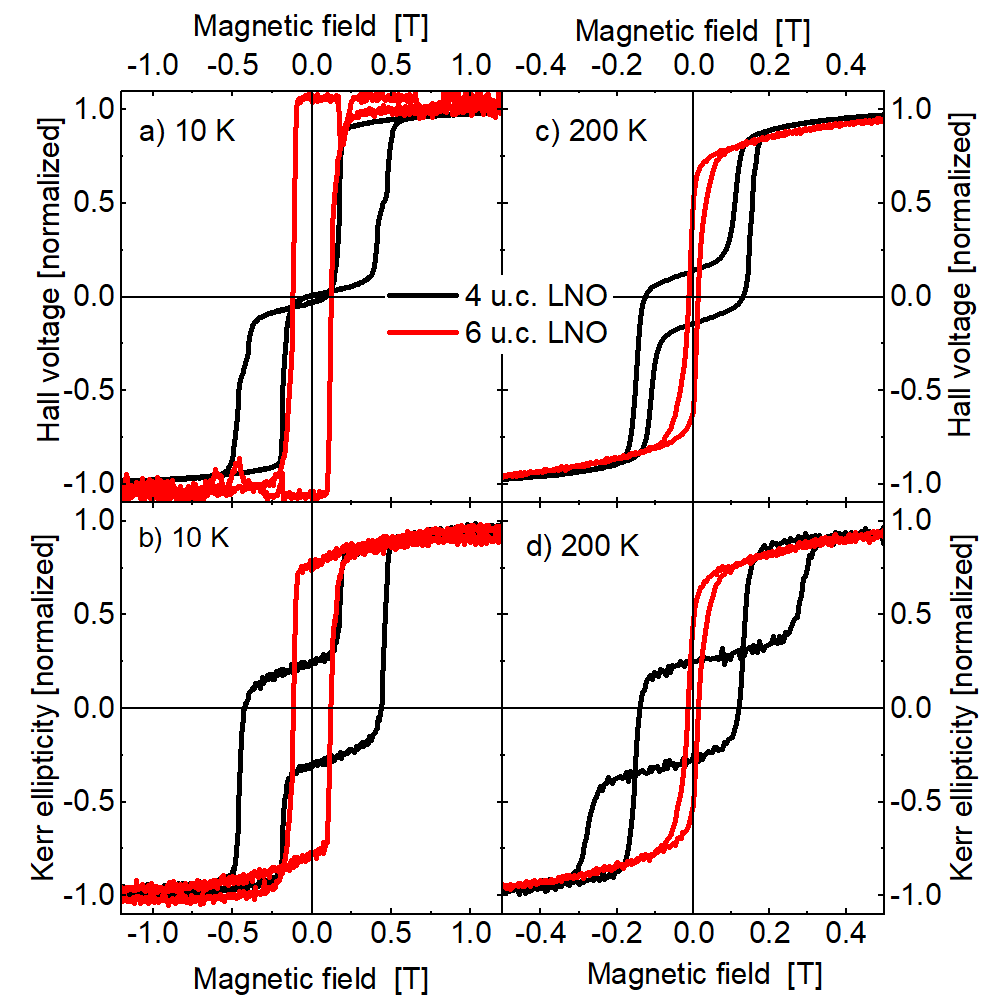}
    \caption{Hysteresis loops of the trilayer samples: (a) Normalized (anomalous) Hall voltage loops, respectively, of samples with a 4 uc thick LNO spacer (black) and a 6 uc thick LNO spacer (red) at 10 K. (b) Kerr ellipticity loops of the same trilayer at 10 K. (c) and (d) The same measurements as in (a) and (b), recorded at 200 K. The magnetic field was applied perpendicular to the sample surface.}
    \label{fig2}
\end{figure}

\section{Sample design and  Methods}
The samples were fabricated using by pulsed-laser deposition (PLD), assisted by reflection high-energy electron diffraction (RHEED), as described in detail in previous publications \cite{Schoepf2023a, Schoepf2023b}. Trilayer samples with two LSMRO layers (about 8 nm thick) separated by a LNO spacer of either 4 unit cells (uc) or 6 uc thickness, as shown in Fig. \ref{fig1}a, and multilayers with five LSMRO layers (about 8 nm thick) seperated by a 4 uc four LNO spacers were grown on LSAT(100) substrates (these substrates were square shaped, measuring either 4 mm or 5 mm in size). Prior to PLD, the LSAT substrates were annealed at 1000 °C in air for 2 hours.

A homemade setup was used  for simultaneous polar magneto-optic Kerr effect (MOKE) and magnetotransport measurements. MOKE measurements utilized a double modulation technique with an optical chopper and a photoelastic modulator (PEM100, Hinds Instruments). Lightsources included a A Xe-lamp with a Jobin Yvon monochromator and a 405 nm Cobolt laser. Magnetotransport measurements were performed using the van der Pauw geometry, with electrical contacts made by gluing copper wires to the sample corners using silver paint. SQUID magnetometry was carried out with a commercially available SQUID magnetometer (MPMS-XL, Quantum Design inc.). Scanning transmission electron microscopy (STEM) of cross-section specimens was conducted with a JEOL 2200FS TEM, equipped with double \textit{C$_s$} (spherical aberration) correctors and operated at 200 keV. STEM specimens were prepared by focused ion beam (JEOL JIB-4700F) milling with Ga ion source. A 1.5 $\mu$m Pt layer was deposited on the surface prior to milling to mitigate ion beam damage.\\

X-ray absorption spectroscopy (XAS) and X-ray circular magnetic dichroism (XMCD) measurements were performed at the 4.0.2 beamline at the Advanced Light Source. Absorption spectra were recorded at the Mn-\textit{L} and Ni-\textit{L} edges using total electron yield (TEY) and luminescence yield (LY) detection modes, at 80 K. LY measurements were particulalry effective for probing the buried LNO layers compared to the the surface-sensitive TEY measurements.

\section{Results and discussion}

The microstructure of the epitaxial heterostructures was investigated by RHEED during growth, which enabled also accurate control over the layer thickness, and post-growth analysis was conducted via atomic force microscopy and STEM. High-angle annular dark field (HAADF)-STEM micrographs of a cross-section specimen of a trilayer sample with a 4 uc thick LNO spacer are shown in Fig. \ref{fig1} b, c. The spacer layer is continuous, with no pinholes or significant  deviations from the nominal 4 uc thickness. Additional microstructural investigations are provided in the \textbf{supplemental material} \cite{Supplemental_Material}. 

In order to assess the type of coupling between LSMRO layers in the trilayer samples, full MOKE (ellipticity and Kerr rotation) and Hall effect measurements were performed simultaneously at temperatures ranging from 10 K to 200 K in a perpendicular magnetic field, as the LSMRO layers exhibit PMA. Data demonstrating the PMA of reference samples, including a single 8 nm thick LSMRO layer and a bilayer with 6 uc LNO on an 8 nm thick LSMRO layer are discussed in the \textbf{supplemental material} \cite{Supplemental_Material}. Figure \ref{fig2} summarizes the normalized Hall voltage and Kerr ellipticity loops of the two type of trilayers, with 4 uc and 6 uc LNO spacers, measured at 10 K and 200 K. These spacer thicknesses were chosen based on expectation that 4 uc LNO would result in AFM coupling and 6 uc LNO would result in FM coupling, similar to La$_{0.67}$Ba$_{0.33}$MnO$_3$-LaNiO$_3$ heterstructures with in-plane anisotropy \cite{Nikolaev2000}. For the Hall voltage loops, the normal Hall effect contribution was subtracted, leaving only the anomalous Hall effect, which is proportional to the sample´s perpendicular magnetization. For the 6 uc LNO spacer, a single loop with a sharp magnetization reversal (red loops) is observed, indicating FM coupling and simultaneous magnetization reversal in the two LSMRO layers. In contrast, the trilayer with a 4 uc LNO spacer shows hysteresis loops (black loops in Fig. \ref{fig2}) characteristic of AFM-coupled layers \cite{Saito2021, Darwin2024}: a sharp step occurs at positive fields of about 0.2 T at 10 K and at 0.13 T at 200 K when decreasing the field from positive saturation to zero. The AFM exchange field is about 0.35 T at 10 K and 0.15 T at 200 K. In other oxide systems, AFM interlayer coupling has been achieved between two La$_{0.67}$Ca$_{0.33}$MnO$_3$ layers with in-plane magnetic anisotropy separated by insulating CaRu$_{0.5}$Ti$_{0.5}$O$_3$ \cite{Jin2023}. However, for this multilayer, the insulating nature of the spacer material suggests that tunneling, rather then \textit{RKKY}, likely dominates the interlayer coupling. The magnetization loops of the samples studied by Jin \textit{et al}. exhibited similar behaviour as the Hall and MOKE loops of our trilayer samples, as summarized in Fig. \ref{fig2}. The AFM exchange field for the La$_{0.67}$Ca$_{0.33}$MnO$_3$/ CaRu$_{0.5}$Ti$_{0.5}$O$_3$ heterostructures in this study was about 30 mT at 20 K.

\begin{figure}[ht]
    \centering
    \includegraphics[width=0.48\textwidth]{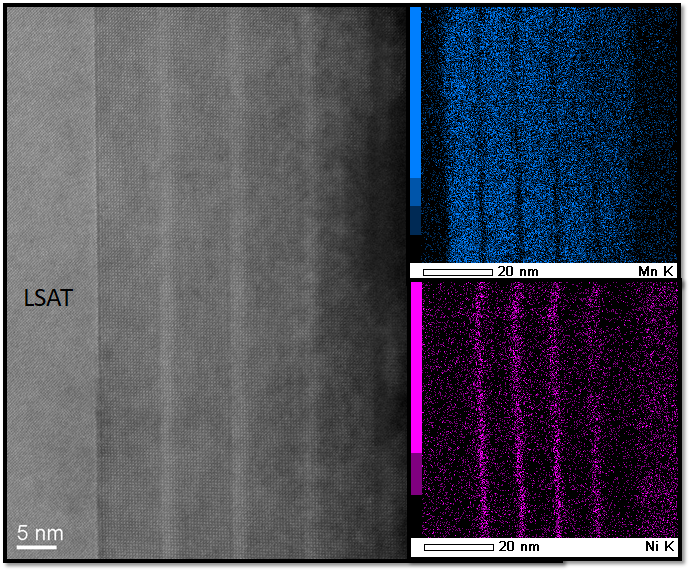}
    \caption{High magnification HAADF-STEM image of the cross section of a multilayer with 4 uc thick LNO spacers between five 8 nm thick LSMRO layers. The images on the right are EDX elemental maps, showing the distribution of Mn (top; blue color) and Ni (bottom; magenta color).} 
    \label{fig3}
\end{figure}

Multilayers with five LSMRO layers and 4 uc thick LNO spacers, made under the same conditions, were studied as well. For an odd number of LSMRO layers it is expected that a finite net magnetization exists at remanence even for AFM interlayer coupling \cite{Darwin2024}. The microstructure of the multilayer wass investigated by STEM and energy dispersive X-ray spectroscopy (EDX), with a summary shown in Fig. \ref{fig3}. The layers, particularly the ultrathin LNO layers, exhibit uniform thickness and sharp interfaces. EDX elemental maps confirm that the Ni signal is confined within the 4 uc thick spacers.

\begin{figure}[ht]
    \centering    \includegraphics[width=0.5\textwidth]{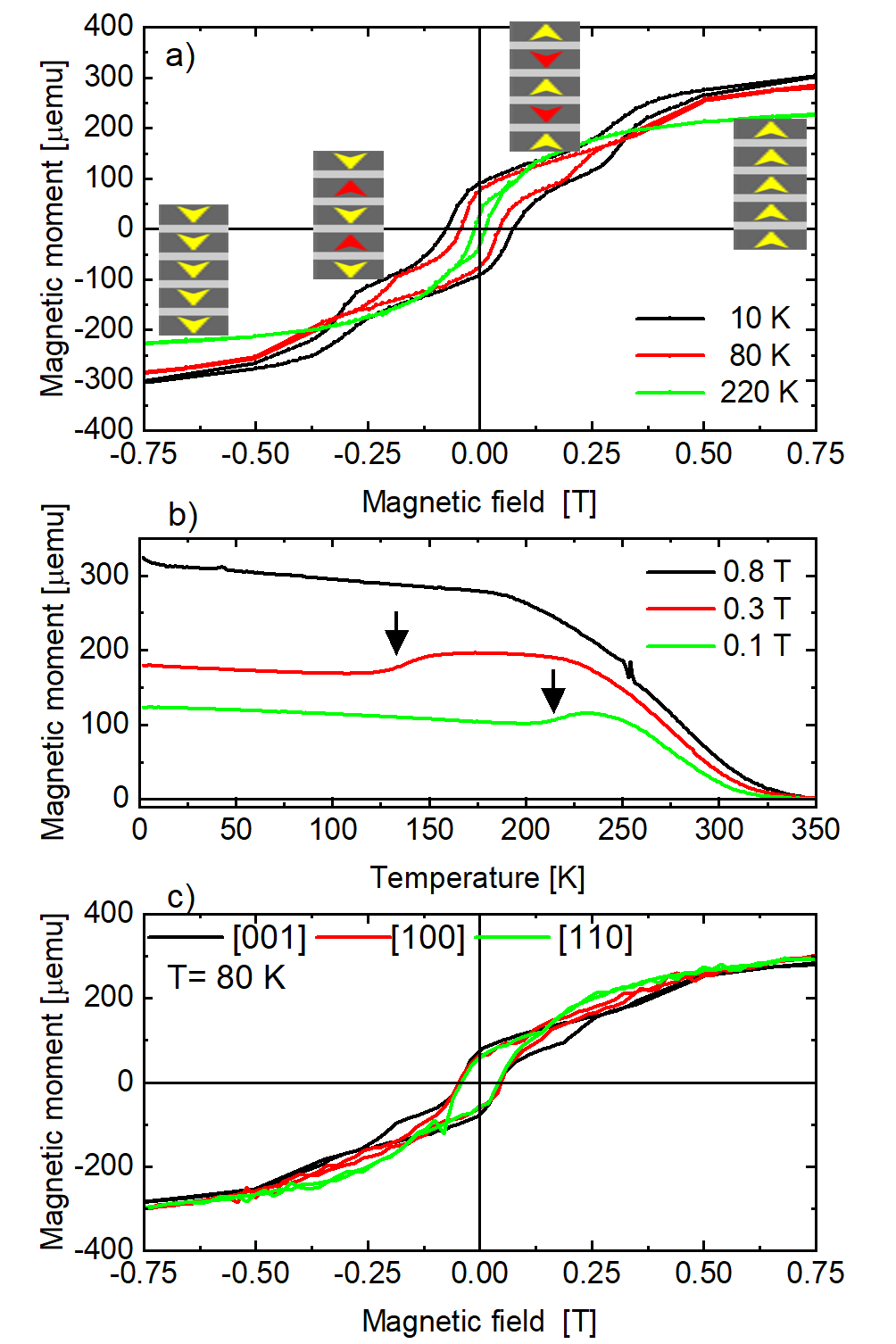}
    \caption{SQUID magnetometry measurements of a multilayer with five LSMRO layers and 4 uc thick LNO spacers. (a) Full loops of the magnetization reversal as a function of applied field at 10 K, 80 K and 220 K, with the field perpendicular to the sample. The schematics of the field-reversal of the perpendicular component of the magnetization in the LSMRO layers are overlaid on the graph. (b) Magnetization as a function of temperature measured by field cooling, for different values of magnetic field along the direction perpendicular to the sample. The black arrows mark the anomalous behavior arising from the AFM-coupling between the LSMRO layers. (c) Magnetization loops for the field applied along different cubic axes, with $[001]$ perpendicular to the sample, at 80 K.} 
    \label{fig4}
\end{figure}

SQUID magnetometry measurements of a multilayer revealed that the effective magnetic anisotropy changes relative to the trilayer samples and that the LNO layers contribute to the magnetization. Figure \ref{fig4} a summarizes the magnetometry measurements. Magnetic hysteresis loops for perpendicular field as a function of temperature are plotted in Fig. \ref{fig4}a. All loops are corrected by subtracting the contribution of the LSAT substrate, which is diamagnetic and has a significant paramagnetic signal at low temperatures (see \textbf{supplemental material})  \cite{Supplemental_Material}. The loop shapes indicate that AFM-coupling exists between the LSMRO layers, as magnetization reversal starts at positive fields when the field is reduced from positive saturation. The presence of AFM-coupling is further supported by the behavior of the magnetization curves measured under field cooling conditions: for lower applied fields (0.1 to 0.3 T, similar to the AFM exchange field), the curves exhibit an anomalous dip, with its position depending on field value (see Fig. \ref{fig4}b, where the dip is marked by black arrows). The dip is suppressed in the M(T) curve measured at 0.8 T, as this field value exceeds the AFM exchange field values at all temperatures down to 10 K. Similar M(T) behavior has been observed for antiferromagnetically coupled La$_{0.67}$Ca$_{0.33}$MnO$_3$/CaRu$_{1/2}$Ti$_{1/2}$O$_3$ multilayers \cite{Chen2017}. The mechanism of layered-resolved magnetization reversal discussed in Ref. \cite{Chen2017, Darwin2024} is plausible for the LSMRO/LNO multilayers as well. However, the loops for the multilayer with five LSMRO layers appear to exhibit only three reversal steps. Analyzing magnetization reversal from M(H) loops alone, without the magnetic domain formation imaging, is more complex for the perpendicularly magnetized multilayers than for those with in-plane anisotropy, as discussed in Ref. \cite{Chen2017}. The hysteresis loop shape is strongly influenced by domain formation, such as stripe domains, due to the strong demagnetization effects present in the PMA situation \cite{Boehm2019}. Therefore, we propose tentatively that the magnetization reverses similarly to the mechanism described in Ref. \cite{Ziese2010} for La$_{0.7}$Sr$_{0.3}$MnO$_3$/SrRuO$_3$ superlattices at 100 K. Looking at the magnetization loop at 10 K (Fig. \ref{fig4}a), the first marked drop in magnetization (at about 0.35 T) corresponds to the onset of magnetization reversal due to the AFM coupling between the LSMRO layers. The second and forth LSMRO layers that reverse their magnetization in positive fields. At fields below about 0.25 T, these two layers reverse their perpendicular magnetization component opposite to the direction of the applied magnetic field direction. A minor loop starting from saturation in high positive field and reversing at 0.2 T shows a shift in the positive direction, consistent with AFM coupling (see \textbf{supplemental material}) \cite{Supplemental_Material}.
In negative fields, three LSMRO layers reverse their magnetization to align with the field, while the other two reverse oppositely due to AFM coupling. Only at negative fields above -0.25 T will these two layers will also reverse to align with the field. The schematics of the magnetization reversal is overlaid in Fig. \ref{fig4}a.  Further evidence for this reversal mechanism is provided by XMCD measurements, discussed below. The effective magnetic anisotropy of the multilayer changed significantly compared to the single LSMRO layer and trilayer samples, which have strong PMA. Similar magnetization hysteresis loop behaviour has been observed for van der Waals ferromagnetic Fe$_3$GaTe$_2$ flakes, five monolayers  thick \cite{Wang2024}. Figure \ref{fig4}c shows the magnetization loops as a function of field direction measured at 80 K. Loops were measured with perpendicular field (along the [001] direction) and with in-plane field along the sample edge ([100] direction) or parallel to the sample diagonal ([110] direction). The comparison of the loops indicates that the effective magnetic anisotropy is no longer dominantly perpendicular to the multilayer surface, as the three loops have similar coercive fields and saturation magnetization, and only slightly different saturation fields. Field cooling M(T) curves measured in three different field directions confirm this statement (data shown in the \textbf{supplemental material}) \cite{Supplemental_Material}. This apparent change in effective magnetic anisotropy, impacting the squareness of the magnetization loops, is partly due to the increased stack thickness (relative to trilayer samples), resulting in a larger contribution of dipolar interactions. The dipolar interactions favor FM coupling between layers and the formation of magnetic stripe domains and/or in-plane reorientation of the magnetization \cite{Hellwig2007, Darwin2024}. 

\begin{figure}[ht]
    \centering
    \includegraphics[width=0.49\textwidth]{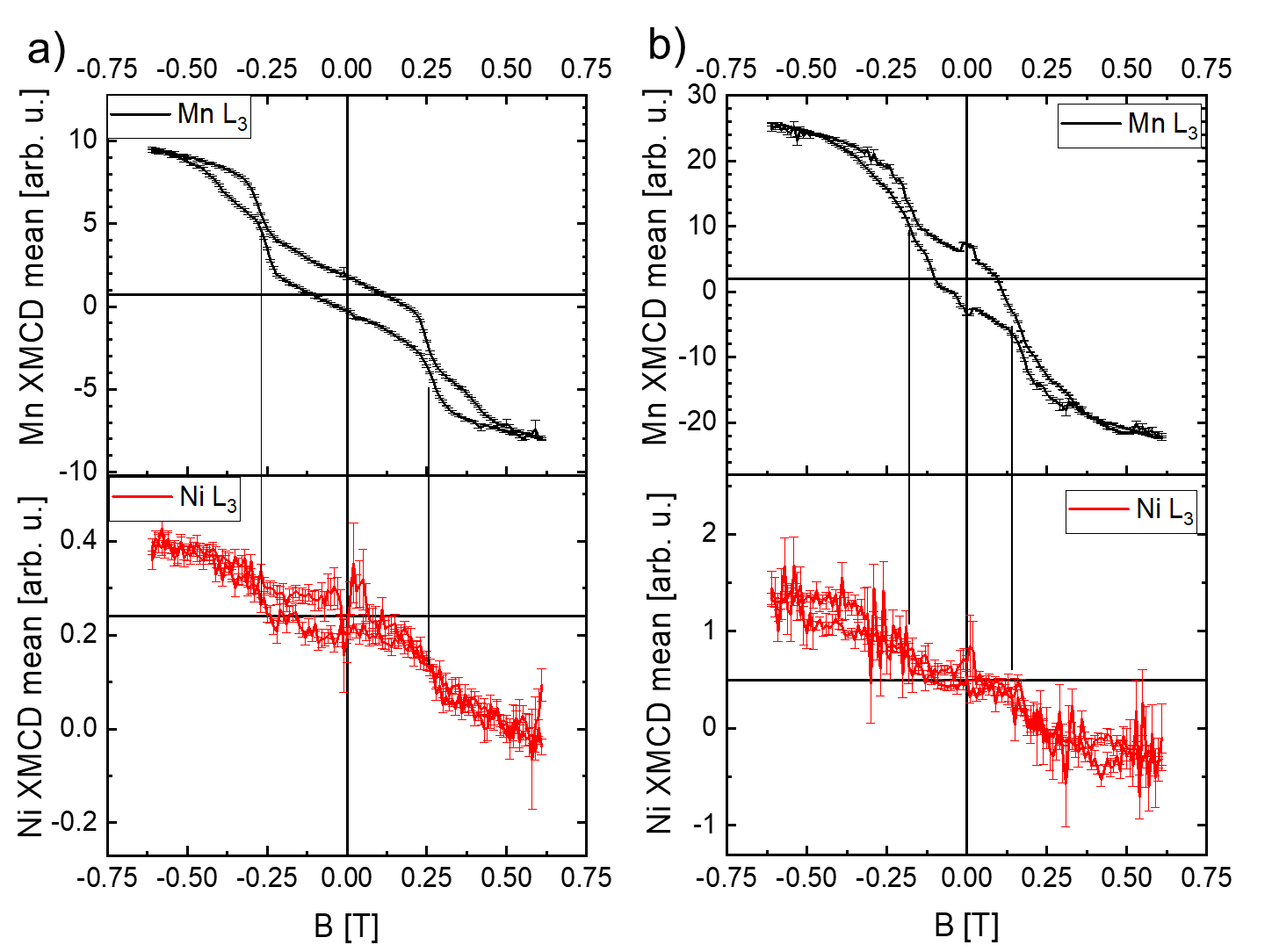}
    \caption{XMCD investigations of a multilayer with five LSMRO layers and 4 uc thick LNO spacers (at 80 K). Loops for Mn \textit{L}-edge (black, top) and for Ni \textit{L}-edge (red, bottom), measured (a) with perpendicular magnetic field and (b) with in-plane magnetic field, in LY conditions.} 
    \label{fig5}
   
\end{figure}

XMCD measurements at the Mn \textit{L}-edge and the Ni \textit{L}-edge were performed in LY-detection mode to better probe the buried LNO layers better \cite{Emori2024}. XMCD is determined from the difference in XAS measurements for right and left circularly polarized incident light. Figure \ref{fig5} shows hysteresis loops measured for the Mn \textit{L}-edge and Ni \textit{L}-edge, for perpendicular and in-plane fields, at 80 K. The in-plane Mn signal is significantly stronger, indicating in-plane canting of the magnetization of the LSMRO layers, in qualitative agreement with the SQUID magnetometry data. However the SQUID loops for in-plane and out-of-plane geometries showed similar magnitudes of the magnetic moment (see Fig.  \ref{fig4}c). Differences between the SQUID magnetometry, which measures the whole sample, and XMCD experiments, which are mainly sensitive to the top-most layers of the sample, can explain discrepancies between the two types of loops. Nonetheless, the shape of the XMCD Mn loops resembles well that of SQUID loops. Additional differences may arise from contributions in the magnetometry loops from the LSAT substrate and from the four LNO layers. For a comparison of SQUID and XMCD normalized loops, see the \textbf{supplemental material} \cite{Supplemental_Material}. Both the perpendicular and in-plane field Mn loops show a sharp initial drop where AFM coupling induces the first switching of two layers (marked by the vertical lines in Fig. \ref{fig5}), followed by an almost linear decrease as the field is reduced to zero and changes its polarity. AFM coupling is overcome in higher fields, leading to a sharp step after which the magnetization again reaches saturation. The linear part strongly indicates the presence of laterally growing stripe domains in the LSMRO layers. This is expected for the thicker multilayer, contrasting with the thinner trilayer, where dipolar interactions favor stripe domains \cite{Hellwig2011}. Stripe domains forming in perpendicular fields have been observed in 30-40 nm thick LSMRO films grown on LSAT substrates \cite{Nakamura2018,phdthesis_Lena}. The magnetic domains in our AFM-coupled LSMRO multilayers are likely more complex than we have assumed \cite{Hellwig2007,Kiselev2007} and require further imaging investigations in perpendicular fields, such as cryogenic magnetic force microscopy. \\

 Regarding the LNO spacer layers, there is a sizeable but weak XMCD Ni-signal indicating magnetic ordering in the LNO layers. The Ni loops suggest that the magnetization of the LNO layers is coupled and follows the magnetization reversal of the LSMRO layers. In La$_{2/3}$Ca$_{1/3}$MnO$_3$/LaNiO$_3$ superlattices grown on SrTiO$_3$ substrates with in-plane magnetic anisotropy, emerging magnetism in LNO was inferred through an exchange bias mechanism at the interfaces observed by X-ray resonant magnetic reflectivity. Soltan \textit{et al}. found non-symmetric interfaces induced magnetization profiles in LNO and LCMO, related to a periodic complex charge and spin superstructure \cite{Soltan2023}.  XMCD investigations of LaMnO$_3$/LaNiO$_3$ superlattices with ultrathin layers (4-7 uc thick individual layers) and in-plane magnetic anisotropy also showed magnetic ordering of the nickelate layers, with Ni magnetic moments being ferromagnetically coupled to Mn magnetic moments \cite{Piamonteze2015}.

\section{Summary}
Here, the interlayer coupling between epitaxial La$_{0.7}$Sr$_{0.3}$Mn$_{0.95}$Ru$_{0.05}$O$_3$ layers with perpendicular magnetic anisotropy, separated by nominally metallic paramagnetic LaNiO$_3$ ultrathin layers, was investigated. Using MOKE, anomalous Hall effect investigations, and SQUID magnetometry, it was found that LaNiO$_3$ serves as a suitable spacer for tuning the interlayer coupling from antiferromagnetic to ferromagnetic, by varying the spacer thickness. The coupling mechanism is of the \textit{RKKY}-type, as discussed by prior studies \cite{Nikolaev2000, Ohsawa2005}. Furthermore,  multilayers with five La$_{0.7}$Sr$_{0.3}$Mn$_{0.95}$Ru$_{0.05}$O$_3$ layers separated by 4 uc thick LaNiO$_3$ spacers were fabricated, showing that the antiferromagnetic interlayer coupling persisted. The magnetization reversal mechanism in the five La$_{0.7}$Sr$_{0.3}$Mn$_{0.95}$Ru$_{0.05}$O$_3$ layers between opposite saturated polarities was proposed, with observations indicating that the effective magnetic anisotropy in the multilayer canted more strongly towards to the in-plane direction, likely due to increased interactions. XMCD investigations revealed that the 4 uc thick LaNiO$_3$ layers in the multilayer were magnetically ordered at 80 K, and their magnetization reversal followed that of the La$_{0.7}$Sr$_{0.3}$Mn$_{0.95}$Ru$_{0.05}$O$_3$ layers. This demonstrates control over the interlayer coupling between manganite layers with perpendicular magnetic anisotropy, achieving strong antiferromagnetic coupling relevant for spintronic applications, such as spin-valve devices. Future work will focus to LSMRO/LNO multilayers, exploring the dependence on thickness and the number of LSMRO layers, as well as the study of magnetic domains in antiferromagnetically-coupled multilayers. These findings open up prospects for designing synthetically antiferromagnetic skyrmion bubbles in ferromagnetic oxide epitaxial multilayers, presenting exciting opportunities for advanced spintronic applications.

\begin{acknowledgments}
ILV and JS wish to acknowledge the financial support from German Research Foundation (DFG) in the framework of CRC1238 (project number 277146847, subproject A01). DK and XZ acknowledge support by the National Science Foundation under Grant No.
NSF DMR1751455. This research used resources of the Advanced Light Source, which is a DOE Office of Science User Facility under Contract No. DE-AC02-05CH11231. STEM investigations were conducted at the OtaNano-Nanomicroscopy Center (OtaNano-NMC), supported by Aalto University in Finland.

\end{acknowledgments}

%\appendix

%\section{Appendixes}

\bibliography{RuLSMO_LNO_interlayer_coupling}% Produces the bibliography via BibTeX.

\end{document}